\documentclass[aps,prl,showpacs,twocolumn,amsmath,amssymb,superscriptaddress,footinbib]{revtex4}

\usepackage[english]{babel}
\usepackage{latexsym}
\usepackage{graphics}
\usepackage{epsfig}
\usepackage{amsfonts}
\usepackage{amssymb}
\usepackage{amsmath}

\begin{document}

\title{Photonic Josephson effect, phase transitions, and chaos in optomechanical systems}


\author{Jonas Larson}
\email{jolarson@fysik.su.se} \affiliation{Department of Physics,
Stockholm University, AlbaNova University Center, Roslagstullsbacken 21, 106 91 Stockholm,
Sweden}

\author{Mats Horsdal} \affiliation{Institute for Theoretical Physics, Leipzig University, D-04009  
Leipzig, Germany}
\affiliation{NORDITA, Roslagstullsbacken 23, 106 91 Stockholm,
Sweden}

\date{\today}

\begin{abstract}
A photonic analog of the Josephson effect is analyzed for a system formed by a partly transparent mechanical membrane dividing an optical cavity into two halves. Photons tunneling between the two sub-cavities constitute the coherent Jospehson current. The force acting upon the membrane due to the light pressure induces a nonlinearity which results in a rich dynamical structure. For example, contrary to standard bosonic Josephson systems, we encounter chaos. By means of a mean-field approach we identify the various regimes and corresponding phase diagram. At the short time scale, chaos is demonstrated to prevent regular self-trapping, while for longer times a dissipation induced self-trapping effect is possible.    
\end{abstract}

\pacs{42.50.Pq, 03.75.Lm, 05.45.Mt}
\maketitle

{\it Introduction}. -- The Jospehson effect describes macroscopic tunneling made possible due to inherent coherence of the underlying quantum many-body state. The resulting Josephson oscillations, however first discussed for weakly coupled superconductors~\cite{je1}, has turned out to be relevant for a wide range of physical systems like, Bose-Einstein condensates in double-well traps~\cite{becdw1}, spinor condensates~\cite{spinbec}, graphene and Hall systems~\cite{hall}, and in nonlinear optics~\cite{pje3}. Lately, the possibility to observe the Josephson effect for photons in systems of coupled high $Q$ optical cavities has been discussed~\cite{pje1,pje2}. Together with related schemes containing arrays of coupled cavities~\cite{hartman} or single cavities filled with a Kerr medium~\cite{polariton}, these proposals pave the way for genuine quantum many-body models to be studied by means of photons. 

Nonlinearity is a basic ingredient for Josephson oscillations; in Ref.~\cite{pje1} a three cavity setup is investigated and nonlinearity is accomplished letting the photons interact dispersively with matter in the middle tunneling cavity, while in~\cite{pje2} only two cavities are coupled and the nonlinearity stems from a gas of ultracold atoms trapped in each cavity. In this work we consider the photonic analog of Josephson
oscillations in an optomechanical cavity system that has been thoroughly analyzed experimentally in the recent past~\cite{harris,shuttle}. Nonlinearity in optomechanical systems is an outcome of the interaction between photons and the mechanical mirror/membrane, and is thereby different from the above schemes which utilize atoms or Kerr mediums in order to induce the effective interaction between photons. Nowadays, the mechanical oscillator can be cooled down close to its quantum ground state~\cite{cool} making these systems promising alternative candidates for observing many-body quantum effects~\cite{optorev}. In our version of the photonic Josephson effect, the mechanical oscillator adds an additional degree of freedom to the dynamics driven by the photon modes, and hence opens up for novel phenomena absent in more regular Josephson systems. We will indeed demonstrate that the system at hand possesses a plethora of phenomena like self-trapping (i.e. localization of photons), bistability, and dynamical phase transitions between chaotic and regular dynamics. We note that the back-action, here induced by the interaction between the membrane and the photon modes, that occurs when the extra degrees of freedom are considered has not been taken into account in~\cite{pje1,pje2} and one expects that those analyses would therefore not predict structures as chaos and attractors.

\begin{figure}[h]
\centerline{\includegraphics[width=5cm]{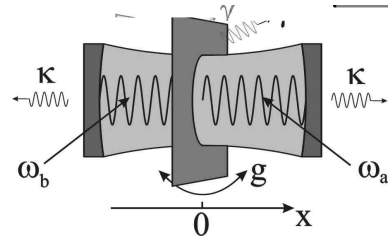}}
\caption{Model system, a partly reflective membrane divides a
Fabry-Perot cavity into two parts $a$ and $b$. With a rate $g$, photons can tunnel between the two parts. At the same time, the membrane is displaced from its equilibrium position ($x=0$) due to the radiation pressure imposed by the photons scattered upon it, and thereby the two effective photon frequencies $\omega_a$ and $\omega_b$ are simultaneously shifted. Photon losses through the two end mirrors are marked by $\kappa$, and phonon decay of the membrane by $\gamma$.} \label{fig1}
\end{figure}

{\it Model system and effective equations}. -- The cavity setup is depicted in Fig.~\ref{fig1}. A thin membrane separates an optical resonator into two halves, where photons can tunnel from one side to the
other~\cite{harris,shuttle}. The optical force acting on the membrane induces a nonlinearity; the position of the membrane determines the photon frequencies which directly affect the Josephson oscillations. By considering an optomechanical system with two fixed end mirrors, the finesse of the cavity can be greatly enhanced in comparison to a two mirror cavity with one fixed and one movable end mirror. In the regime where the photon modes
follow adiabatically the membrane displacement~\cite{law}, non-zero transmitivity of the middle mirror brings out constant shifts of the two photon frequencies as well as optomechanical couplings of higher order in the membrane displacement than the linear one which would be the case of a perfectly reflecting membrane~\cite{meystre}. We will assume small membrane displacements, i.e. much smaller than the photon wavelengths, and the mirror to be positioned to render a linear coupling in $x$~\cite{shuttle}.

In the following we work with dimensionless variables, letting the membrane excitation energy, $\hbar\omega_m$, set a characteristic energy scale, i.e. giving a length scale $l=\sqrt{\hbar/m\omega_m}$ with $m$ the membrane mass. Denoting by $\hat{x}$ and $\hat{p}$ the dimensionless position and momentum of the membrane, $\hat{a}$ and $\hat{b}$
($\hat{a}^\dagger$ and $\hat{b}^\dagger$) annihilation (creation) photon operators of the two cavity modes, the Hamiltonian
is~\cite{shuttle}
\begin{equation}\label{ham1}
\begin{array}{lll}
\hat{H}_{om} & = & \displaystyle{\frac{\hat{p}^2}{2}+\frac{\hat{x}^2}{2}+\omega\left(\hat{n}_a+\hat{n}_b\right)}\\ 
& & \displaystyle{+g\left(\hat{a}^\dagger\hat{b}+\hat{b}^\dagger\hat{a}\right)+\lambda\hat{x}\left(\hat{n}_a-\hat{n}_b\right)}.
\end{array}
\end{equation}
Here, we have taken the equilibrium position of the membrane in the absence of photons $\langle\hat{x}\rangle=0$ and assumed the effective frequencies of the two photon modes to be the same and equal to $\omega$, $g$ is the effective tunneling coefficient of photons between the left and right regions, $\lambda$ the effective optomechanical coupling strength, and $\hat{n}_a=\hat{a}^\dagger\hat{a}$ and
$\hat{n}_b=\hat{b}^\dagger\hat{b}$ are the photon number operators of the two modes. The number of photons $\hat{N}_{ph}=\hat{n}_a+\hat{n}_b$ is conserved and we go to an
interaction picture rotating with respect to $\omega\hat{N}_{ph}$. In the new basis, the third term of $\hat{H}_{om}$ is lacking and we are left with only two effective parameters $g$ and $\lambda$.

In the decoupled case, $\lambda=0$, and assuming that initially all photons resides
in one of the modes, as time progresses the photons will tunnel through the membrane showing perfect Rabi oscillations with a frequency $\Omega=g$. This is irrespective of the number
of photons, and hence after half a Rabi period all photons have been swapped to the opposite mode. For $\lambda\neq0$ the dynamics become much more complex. The shift of the membrane is proportional to the imbalance of population between the two modes, and thereby the dynamics is highly nonlinear. As the displacement induces an effective detuning between the two modes, for large displacements the tunneling of photons will be greatly reduced, i.e. self-trapped. 

We analyze the dynamics within a mean-field approach that has proved to accurately describe the experimental Josephson effect in Bose-Einstein condensates~\cite{becdw1}. We write down the Heisenberg equations of motion for the photon inversion $\hat{z}=\left(\hat{n}_a-\hat{n}_b\right)/N_0$, the phase difference
$\hat{\phi}=\arg\left(\hat{a}^\dagger\hat{b}\right)$, the photon loss fraction $\hat{q}=(\hat{n}_a+\hat{n}_b)/N_0$, and $\hat{p}$ and $\hat{x}$, and treat them as $c$-numbers to obtain a closed set of coupled equations. $N_0$ is the number of photons at time $t=0$. This method allows for losses of both the photon and phonon fields to be taken into account, and we therefore have introduced the quantity  $\hat{q}$. The mean-field equations of motion read
\begin{equation}\label{eom}
\begin{array}{lll}
\dot{x} & = & \displaystyle{p-\frac{\gamma}{2}x},\\
\dot{p} & = & \displaystyle{-x-\frac{\gamma}{2}p-\lambda N_0z},\\
\dot{z} & = & \displaystyle{2g\sqrt{q^2-z^2}\sin\phi-\kappa z},\\
\dot{\phi} & = & \displaystyle{2\lambda x-2g\frac{z}{\sqrt{q^2-z^2}}\cos\phi} ,\\
\dot{q} & = & -\kappa q+2\kappa\frac{N_{th}}{N_0},
\end{array}
\end{equation}
where $\kappa$ is the photon decay rate and $\gamma$ the phonon decay rate. The photon reservoir temperatures are determined by the average number of thermal photons $N_{th}$ at the frequency of the two cavity modes, while the phonon reservoir temperature does not enter since we have truncated the equations of motions to contain only $x$ and $p$ and not any higher orders. 

\begin{figure}[h]
\centerline{\includegraphics[width=7cm]{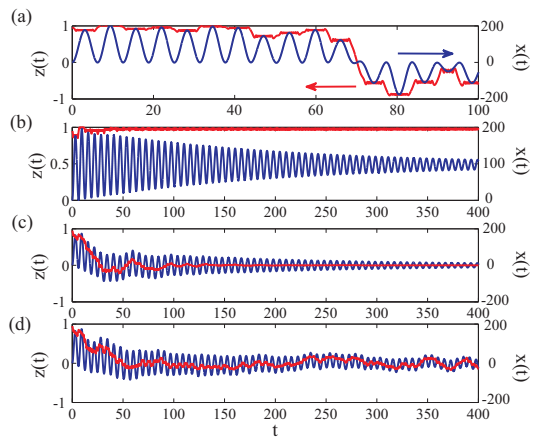}}
\caption{(Color online) Time evolution of the mean-field photon inversion $z(t)$ (red curves) and membrane position $x(t)$ (blue curves). In (a), the system is closed/conservative, $\gamma=\kappa=0$, and the dynamics is chaotic with the inversion making sudden jumps as the membrane approaches $x=0$. In (b), damping of the membrane has been included, $\gamma=0.01$, $\kappa=0$, and the system reaches a self-trapped state (an attractor). In (c), $\gamma=0.01$, $\kappa=0.02$, and $N_{th}=0$, and as photons leak out of the cavity the membrane's displacement vanishes. Finally, in (d) the situation is the same as in (c) but with $N_{th}=200$ causing the chaotic behavior of detuned Josephson oscillations to last throughout the evolution. In all four plots, the dimensionless parameters are $g=0.2$, $\lambda=0.1$, $N_0=1000$, and the initial conditions $x_0=p_0=\phi_0=0$, $z_0=0.95$, and $q=1$. Thus, the parameters are such that the system starts out in the chaotic regime in all examples
 . Other initial conditions show similar behaviors provided the system starts out far from any fixed point or attractor.}
\label{fig3}
\end{figure}

{\it Self-Trapping, bifurcation, and chaos}. -- Several general conclusions can be readily drawn from the set of mean-field equations~(\ref{eom}); the first two equations describe a damped oscillator ``driven'' by the term proportional to $z$, the third and forth equations represent a pendulum with varying length and ``driven'' by $2\lambda x$. In the limit when the shift of the membrane can be approximated as proportional to $z$, one recovers the standard mean-field equations of motion for a Bose-Einstein condensate in a double-well~\cite{becdw1}. However, such an approximation is in general not justified and the set of equation~(\ref{eom}) turn out to possess a dynamical richness not found in regular Bose-Einstein condensate double-wells. The various types of dynamical behavior can be divided in:

{\it (1) Lossless case, $\kappa=\gamma=0$}. In the lossless case the system is conservative in the sense that a volume-element in phase space preserves its volume in time. The solutions evolve on four dimensional tori, which become more and more deformed the stronger the nonlinearity gets in accordance with KAM-theory~\cite{kam}. At $C\equiv g/\lambda^2N_0=1$ there is a pitchfork bifurcation characterizing a second order dynamical phase transition~\cite{chaos}. For $C>1$, there is one center fixed point, $z=x=p=0$ and $\phi=(2n+1)\pi$ ($n$ an integer), and the system shows Josephson like dynamics, i.e. far from the fixed point the photons oscillate between the two modes. For $C<1$, when nonlinearity becomes important, the above fixed point becomes unstable and two additional center fixed points appear, $z=\pm\sqrt{1-C^2},\,x=-\lambda N_0 z,\,p=0,\,\phi=(2n+1)\pi$. In this regime, self-trapping is expected, but is here prevented by chaotic dynamics, as demonstrated in Fig.~\ref{fig3} (a) displaying time evolution of $z$ and $x$~\cite{note}. The membrane shows an oscillatory behavior; when it is displaced far from $x=0$ the effective detuning between the two photon modes is large and $z$ shows detuned Josephson oscillations, when the membrane approach $x=0$ the detuning vanishes and the photon inversion reveals sudden jumps. Such jumps inhibit the system from becoming self-trapped, and moreover, small fluctuations in system parameters or dynamical variables render completely divergent long time evolutions as is characteristic for chaos. We should point out that there is as well an additional fixed point for $\phi=2n\pi$ and $x=p=z=0$.  

{\it (2) Membrane losses, $\gamma\neq0$, $\kappa=0$}. Contrary to the above case, for losses of the membrane the system is dissipative, i.e. a volume element in phase space decreases with time~\cite{chaos}. A similar pitchfork bifurcation as above is found at $C(1+\gamma^2/4)=1$, but this time the fixed points are stable. Demonstrated in Fig.~\ref{fig3} (b), far from the fixed points the system evolution is in general chaotic to start with, but after some time it approaches attractors that encompass the fixed points for which the membrane is locked and $z$ and $\sin(\phi)$ (we consider $\sin(\phi)$ instead of $\phi$ since the system is periodic in $\phi$) evolve on an ellipse-like surface. Physically, the nonzero inversion implies a displacement of the membrane, while a nonzero loss rate $\gamma$ tend to restore the membrane to its equilibrium position $x=0$. Since the photon number is conserved, the membrane attains a state of self-lasing where the light force and the friction balance one another. In this respect the system becomes self-trapped. We note that similar attractors were found for a single mode optomechanical system in the linear regime~\cite{att1}. As long as we do not drive the cavity modes, the asymptotic self-trapped state is {\it a priori} hard to predict because of the chaotic behavior preceding the system's convergence of the attractor. 

{\it (3) Cavity and mirror losses at zero temperature, $\gamma\neq0$, $\kappa\neq0$, $N_{th}=0$}. When the system is not externally driven and all boson modes possess losses, the only fixed point is the one of all modes empty. This stationary solution is reached on a timescale $T>\max\{\gamma^{-1},\kappa^{-1}\}$. Nevertheless, for times $t\ll T$ the dynamics will be determined by the size of $C(1+\gamma^2/4)$ as shown in Fig.~\ref{fig3} (c). The decay rate $\kappa$ has been taken as $\kappa=0.1g$ in agreement with the experimentally relevant work~\cite{shuttle}. Up till $t\sim50-100$, the dynamics is similar to that displayed in (a), i.e. detuned Josephson oscillations interrupted by sudden jumps. In this example $\kappa^{-1}<\gamma^{-1}$ implying that after a period of chaotic evolution, the larger decay of the photon modes causes them to decay into vacuum, while the membrane sustains its oscillations for some longer times.

{\it (4) Cavity and mirror losses at non-zero temperature, $\gamma\neq0$, $\kappa\neq0$, $N_{th}\neq0$}. When all modes couple to their respective reservoirs, an energy source is required in order to prevent decay into the trivial vacuum steady state solution. This could either be  non-zero temperature baths or direct pumping~\cite{pje1}. For $N_{th}\neq0$, the equations exhibit a nonvanishing fix point for the photon loss fraction $q=q_0\equiv 2\frac{N_{th}}{N_0}$. Also this case includes a pitchfork bifurcation. When $D\equiv \frac{\kappa}{2g}>1$ the bifurcation is subcritical; an unstable fix point for $\tilde{C}\equiv C(1+\gamma^2/4)/q_0<1$ splits into two unstable and one stable fix point at $\tilde{C}=1$. When $D<1$, the bifurcation is supercritical and one stable fix point for $\tilde{C}>1$ splits into two stable and one unstable fix point at $\tilde{C}=1$. For this situation, we find as well two saddle node bifurcations at $\tilde{C}=\frac{2D}{1+D^2}<1$, where the two stable fix points in the pitchfork bifurcation meet two outer unstable fix points. Using the proper physical parameters $\kappa$ and $\gamma$ of the examples presented in Fig.~\ref{fig3} (c) and (d), the parameter $D=1$ is exactly at the separation between the two parameter regimes. The chaotic behavior we study manifest itself, however, far from the critical points and the actual value of $D$ is of minor importance in comparison to $\tilde{C}\lessgtr1$.  When the coupling between the membrane and the cavity modes are non-zero, the evolution stays chaotic comprising detuned Josephson oscillations between sudden jumps in the inversion, see Fig.~\ref{fig3} (d). We have also studied the case of $N_{th}=0$ but with pumping of the two photon modes. The dynamics turn out very similar to that encountered for a non-zero temperature bath. 

As is well known, a fixed point analysis does not reveal the full characteristics of dynamical systems~\cite{chaos}. We find that especially for $N_{th}\neq0$, the long-time solutions of~(\ref{eom}) exhibit a rich structure. As the nonlinearity is increased, the long-time solutions cover a series of periodic doublings in terms of Hopf bifurcations~\cite{chaos}, to finally enter into a regime of chaos. In a driven system, this feature should be captured by varying the pump strength, where each Hopf bifurcation render a dynamical second order phase transition.     

{\it Concluding remarks}. -- To summarize, we have proposed the use
of optomechanical systems for the study of a photonic counterpart of the Josephson
effects. The system was shown, on a mean-field level, to display bifurcations and dynamical phase transitions, as well as a chaotic behavior. In this regime, a new phenomenon of reservoir assisted self-trapping was established. Throughout we used dimensionless parameters. Considering experimentally relevant parameters for optical cavities~\cite{shuttle,meystre}, one has typically $10^{-6}<\lambda<10^{-1}$ and $0.2<g<100$. In this work we choose the upper and lower limiting cases of $g$ and $\lambda$ respectively to achieve the strongest possible nonlinearity,  giving a characteristic time scale of one $\mu s$. However, similar results are obtainable with other values on the cost of longer time scales. In real units and using common experimental parameters~\cite{shuttle,meystre}, one finds that the membrane displacement is several orders of magnitude smaller than typical photon wavelengths implying that the linear coupling assumption is justified. The cavity decay rate $\kappa=0.02$ was taken from  Ref.~\cite{shuttle} and should be of experimental relevance. Utilizing these parameters, Fig.~\ref{fig3} demonstrated establishment of both Josephson oscillations as well as the chaotic jumps before the role of dissipation becomes too important. It is worth noting that for typical cavity frequencies, $N_{th}=200$ as in Fig.~\ref{fig3} (d) implies extremely hot photon baths, and it is thereby more likely that external pumping of the cavities is experimentally preferable. We especially note that since the parameter $\tilde{C}\propto N_0^{-1}$, by increasing such cavity pumping $\tilde{C}$ can, in principle, be made arbitrary small implying strong nonlinearity and chaotic dynamics. Consequently, by measuring the intensity of the outgoing field, the characteristics of the evolution (Rabi, Josephson, chaotic) should be easily achievable with current state-of-the-art optomechanical experiments~\cite{harris}. The analysis is purely semi-classical and any reservoir induced fluctuations have been neglected~\cite{walls}. In the large excitation limit this cannot be a shorting as verified by the current phonon lasing experiment~\cite{lasing} (see also~\cite{lznew}). In the more purely quantum regime however, decoherence outings may become more significant for the coherent tunneling process. In~\cite{pje1}, a full quantum analysis for a similar system was presented and it was found that for external pumping the Josephson effects survives despite coupling to surrounding reservoirs. Moreover, most recently a photonic analog of the intrinsic Josephson effect was studied and it was found that even deep in the quantum regime, the Josephson effect survives around 10 full oscillations~\cite{jonasJJ}. The same is supposedly true in the present system, and we have verified numerically that the same type of mean-field dynamical structures are encountered for coherently driven cavities. Beyond the results of the present work, purely quantum effects like squeezing and anti-bunching are most certainly interesting and will be studied in a forthcoming publication.
  
\begin{acknowledgements}
We thank Ralf Eichhorn for helpful discussions. JL acknowledges support from the Swedish research council (VR).
\end{acknowledgements}


\begin{thebibliography}{999}

\bibitem{je1} B. D. Josephson, Phys. Lett. {\bf 1}, 251 (1962); P. L. Anderson and J. W. Rowell, Phys. Rev. Lett. {\bf 10}, 230 (1963).

\bibitem{becdw1} A. Smerzi {\it et al.}, Phys. Rev. Lett. {\bf 79}, 4950 (1997); M. Albiez {\it et al.}, Phys. Rev. Lett. {\bf 95}, 010402 (2005); S. Levy {\it et al.}, Nature {\bf 449}, 579 (2007).

\bibitem{spinbec} J. Williams {\it et al.}, Phys. Rev. A {\bf 59}, R31 (1999).

\bibitem{helium} S. V. Pereverzev {\it et al.}, Nature {\bf 388}, 449 (1997); K. Sukhatme {\it et al.}, Nature {\bf 411}, 280 (2001).

\bibitem{hall} I. B. Spielman {\it et al.}, Phys. Rev. Lett. {\bf 84}, 5808 (2000); M. Titov and C. W. J. Beenakker, Phys. Rev. B {\bf 74}, 041401 (2006).

\bibitem{pje3} R. Khomeriki, J. Leon, and S. Ruffo, Phys. Rev. Lett. {\bf 97}, 143902 (2006).

\bibitem{pje1} D. Gerace {\it et al.}, Nature Phys. {\bf 5}, 281 (2009).

\bibitem{pje2} A.-C. Ji {\it et al.}, Phys. Rev. Lett. {\bf 102}, 023602 (2009).

\bibitem{hartman} M. J. Hartmann {\it et al.}, Laser Photonics Rev. {\bf 2}, 527 (2008).

\bibitem{polariton} J. Pasprzak {\it et al.}, Nature {\bf 443}, 409 (2006).

\bibitem{harris} J. D. Thompson {\it et al.}, Nature {\bf 452}, 72 (2008).

\bibitem{shuttle} G. Heinrich, J. G. E. Harris, and F. Marquardt, Phys. Rev. A {\bf 81}, 011801(R) (2010).

\bibitem{cool} Y. S. Park and H. Wang, Nature Phys. {\bf 5}, 489 (2009); A. Schliesser {\it et al.}, Nature Phys. {\bf 5}, 509 (2009); T. Rocheleau {\it et al.}, Nature {\bf 463}, 72 (2010).

\bibitem{optorev} T. J. Kippenberg and K. J. Vahala, Science {\bf 321}, 1172 (2008); F. Marquardt and S. M. Girvin, Physics {\bf 2}, 40 (2009).









\bibitem{law} C. K. Law, Phys. Rev. A {\bf 51}, 2537 (1995).

\bibitem{meystre} M. Bhattacharya, H. Uys, and P. Meystre, Phys. Rev. A {\bf 77}, 033819 (2008).










\bibitem{kam} R. C. Hilborn, {\it Chaos and Nonlinear Dynamics}, 2nd ed. (Oxford University Press, New York, 1980).

\bibitem{chaos} S. H. Strogatz, {\it Nonlinear Dynamics and Chaos}, (Westview Press, Cambridge, 2000).

\bibitem{note} To solve the differential equations we utilize the Adams-Bashforth-Moulton method especially suited for problems with stringent error tolerances. 

\bibitem{att1} F. Marquardt {\it et al.}, Phys. Rev. Lett. {\bf 96}, 103901 (2006).

\bibitem{walls} D. F. Walls and G. J. Milburn, {\it Quantum Optics}, (Springer Verlag, 2010, Berlin).

\bibitem{lasing} I. S. Grudini {\it et al.}, Phys. Rev. Lett. {\bf 104}, 083901 (2010).

\bibitem{lznew} H. Wu, G. Heinrich, and F. Marquardt, arXiv:1102.1647.

\bibitem{jonasJJ} J. Larson, Phys. Rev. A {\bf 83}, 052103 (2011).

\end{thebibliography}
\end{document}